\documentclass[12pt]{article}
\begin{document}
\begin{center}
\textbf{Consistent Quantum - Classical Interaction and Solution of the 
Measurement Problem in Quantum Mechanics}

\vspace{.2in}
Tulsi Dass \\
\textsl{Chennai Mathematical Institute, Plot No H1, SIPCOT IT Park, 
Padur Post, Siruseri, Tamil Nadu,  603103 ,India} 
\end{center}

\begin{sloppypar}
\vspace{.2in}
\noindent
\textbf{Abstract}

\noindent
Supmech, an algebraic scheme of mechanics integrating noncommutative 
symplectic geometry and noncommutative probabaility, subsumes quantum and 
classical mechanics and permits consistent treatment of  interaction of 
quantum and classical systems. Quantum measurements are treated in this 
framework; the von Neumann reduction rule 
(generally postulated) is \emph{derived} and interpreted in physical terms. 

\end{sloppypar}

\vspace{.15in}
\noindent
PACS classification codes: 03.65.Ta, 03.65.Ca

\vspace{.12in}
\newline   Noncommutative symplectic geometry 

\vspace{1.8in}
\noindent
Tel. Off.  +91-44-3298 3441; \  +91-44-3298 3442

\noindent
E-mail: tulsi@cmi.ac.in

\newpage

\begin{sloppypar}

\noindent
\emph{Yes, indeed \\
It takes shape \\
When observed.}

\vspace{.15in}
\noindent
\textbf{1.} In the theoretical treatment of a measurement on a quantum 
system, one needs to describe the interaction between the measured quantum 
system  and the apparatus which, being generally macroscopic, is classical. 
Absence of a satisfactory formalism to describe such an interaction led the 
theoretitians [1,2] to treat it as a 
purely quantum mechanical problem (all systems being, presumably, quantum 
mechanical). Consider, for simplicity, 
the measurement of a physical quantity represented by a self-adjoint operator 
F having a nondegenerate spectrum with the eigenvalue equations $ F |\psi_j> 
= \lambda_j | \psi_j> (j = 1,2,...).$ The $ | \psi_j>s$ constitute a basis 
in the  Hilbert space $ \mathcal{H}_S$ of the measured system S. To each of 
the eigenvalues $ \lambda_j$ (which are supposed to be the possible outcomes 
of the measurement) corresponds a pointer position $ M_j.$ Treating the 
apparatus quantum mechanically, one associates, with these pointer positions, 
quantum states $ | \mu_j>$ lying in the apparatus Hilbert space 
$ \mathcal{H}_A.$ The coupled system (S+A) has the Hilbert space 
$\mathcal{H} = \mathcal{H}_S \otimes \mathcal{H}_A$. The measurement 
interaction 
is elegantly described [3,4] by a unitary operator U on $\mathcal{H}$ which, 
acting on the initial state $|\psi> \otimes |\mu_0>$ (where $|\mu_0>$ is the 
`ready' state of the apparatus) an appropriate final state. If the initial 
state of S is one of the eigenstates $|\psi_j>$, the final outcome must be 
(assuming that the measurement is ideal) $\lambda_j$ implying $U (|\psi_j> 
\otimes |\mu_0>) = |\psi_j> \otimes |\mu_j>$.  
For S in the initial state $ |\psi> = \sum c_j |\psi_j>$, the final (S+A)-state
must be, by linearity of U, 
\begin{eqnarray}
|\Psi_f> \equiv U [ (\sum_j c_j |\psi_j>) \otimes |\mu_0>] = 
\sum_j c_j [|\psi_j> \otimes |\mu_j>].
\end{eqnarray}
Experimentally, however, one obtains, in each measurement, a definite outcome 
$\lambda_j$ corresponding to the final (S+A)-state $|\psi_j> \otimes |\mu_j>$.
To account for this, von Neumann postulated that, after the operation of the 
measurement interaction, a non-causal process takes place which changes the 
state $ |\Psi_f>$ to the one represented by the density operator
\begin{eqnarray}
\rho_f = \sum_i P_i |\Psi_f><\Psi_f| P_i = 
\sum_j |c_j|^2 [|\psi_j><\psi_j| \otimes |\mu_j><\mu_j|]
\end{eqnarray}
with $P_i = |\psi_i><\psi_i| \otimes I_A$ where $I_A$ is the unit operator on 
$\mathcal{H}_A$. Eq.(2) predicts that, on repeated measurements, the various 
outcomes $\lambda_j$ appear with respective probabilities $ p_j = |c_j|^2$. 
This is in agreement with experiments. 
The measurement problem in quantum mechanics (QM) is that of explaining, 
starting 
with the  initial state as above, the final state (2), without making any 
ad-hoc assumptions.

A serious  attemt to solve this problem within the framework of traditional 
QM, invokes the interaction of the system (S+A) with the environment which 
results in a rapid suppression of the interference terms in the quantity 
$\zeta \equiv |\Psi_f><\Psi_f| - \rho_f$ (environment -induced decoherence 
[5]). 
A critical evaluation of this approach, however, shows [6,7,4] that it does 
not really solve the problem. In the decoherence formalism, the reduced 
density operator of (S + A) is obtained by 
taking trace (over the environmnt $\mathcal{E}$) of the density operator of 
(S + A + $\mathcal{E}$). Zurek [8] interprets this as ignoring the 
uncontrolled and unmeasured degrees of freedom. This is supposed to be taken 
as similar to the procedure of deriving the probability 1/2 for `heads' as 
well as `tails' in the experiment of tossing a fair coin by averaging over the 
uncontrolled and unmeasured degrees of freedom of the environment of the coin. 

The two procedures are, however, substantially different [6]. In the coin toss 
experiment, when, ignoring the environment, we claim that the probability of 
getting `heads' in a particular toss of the coin is 1/2, we can also claim 
that we do, in fact, get \emph{either} `heads' \emph{or} `tails' on each 
particular toss. A definite outcome can be predicted if we take into 
consideration appropriate environmental parameters and details of initial 
conditions of the throw. In the case of a quantum measurement (as treated in 
the decoherence formalism), however, we  cannot claim that, taking the 
environment into consideration, a definite outcome of the experiment will be 
predicted. In fact, taking the environment into account will give us back a 
troublesome equation of the form of Eq.(1) [with A replaced by (A + 
$\mathcal{E}$)] which is obtained in a von Neumann type treatment of the system 
(S + A + $\mathcal{E}$). 

The problem really lies with the ad-hoc nature of the professedly 
quantum theoretic treatment of the apparatus  and not treating it 
 properly as a system. We shall see below that a consistent formalism describing 
the interaction between a quantum and a classical system can be developed 
and that it is adequate to treat the apparatus  (properly as a \emph{system}) 
classically in such a formalism.

\noindent
\textbf{2.} A formalism of the above sort is provided by supmech [9] 
(which is a refinement and extension of the formalism presented in section 
IV of Ref. [10]), a very 
general sheme of mechanics which accomodates (autonomously developed) 
 QM and classical mechanics as subdisciplines and facilitates a 
transparent treatment of quantum-classical correspondence. It is based on 
the framework of non-commutative geometry (NCG) evolved by Dubois Violette and 
coworkers [11-13]. The underlying idea is that differential forms (covariant 
antisymmetric tensor fields) on a manifold M and objects and operations 
related to them can 
be defined in purely algebraic terms, taking the algebra of smooth functions 
on M [ the commutative algebra $C^{\infty}(M)$ with (fg)(x) = f(x)g(x) = 
(gf)(x)]as the central object (instead of the manifold M). Most of this 
development does not use commutativity of $C^{\infty}(M)$ and, therefore, 
permits an extension to a general (complex, associative, not necessarily 
commutative) algebra $\mathcal{A}$; in particular, one can define 
differential form like objects on $\mathcal{A}$ which are referred to as 
non-commutative differential forms. The traditional differential forms are  
special cases of these for $\mathcal{A} = C^{\infty}(M).$ 

The central object in a proper geometrical treatment of classical Hamiltonian 
mechanics 
is the symplectic form (a  covariant antisymmetric second rank tensor field 
on the phase space $\Gamma$ of of the system which, considered as a matrix, 
is nonsingular at each point of $\Gamma$). The inverse of this tensor 
(which is a contravariant second rank antisymmetric tensor) is used in the 
construction of Poisson brackets (PBs). [ Combining the tradional phase space 
variables $q^j, p^j$(j=1,..,n) into $u^a$ (a = 1,..,2n), we can write the PB 
as $ \{ f,g \} = \omega^{ab}(u)\frac{\partial f}{\partial u^a} 
\frac{\partial g}{\partial u^b}$ where $\omega^{ab}(u) = - \omega^{ba}(u)$ 
and and $det[\omega^{ab}(u)] \neq 0 $ at all points u in $\Gamma$.

In NCG one can define (on an algebra $\mathcal{A}$ as above) a non-commutative 
symplectic form and use it to construct PBs $\{A, B \}$ (where A,B are 
elements of $\mathcal{A}$) which have the usual 
properties  of bilinearity, antisymmetry, Jacobi identity and the Leibnitz rule. The traditional PBs are the special case with $\mathcal{A} = C^{\infty}
(\Gamma)$ and symplectic form $\omega_{ab}$
(the inverse of the tensor field $\omega^{ab}$ 
above). The simplest example of a  noncommutative algebra with a PB defined on 
it  is the algebra $M_n(C)$ (the algebra of complex $n \times n$ matrices) for 
$n\geq 2$ with the commutator [A,B] as PB.

Supmech, which is essentially non-commutative Hamiltonian mechanics, is a 
scheme employing observables and states. It associates, with a system S, 
an algebra $\mathcal{A}$ of the above sort which has a unit element (denoted as I) and 
is a *-algebra in the sense that an antilinear *-operation is defined on it 
which satisfies the relation $ (AB)^* = B^* A^* $. Its Hermitian elements 
(i.e. those elements A with $A^* = A$) are observables of S. A state of 
S is a complex linear functional $\phi$ on $\mathcal{A}$ which is positive 
[i.e. $\phi(A^*A) \geq 0$ for all A in $\mathcal{A}$] and normalised 
[i.e. $\phi (I) = 1$]. The value $ \phi (A)$ of an observable A in a state 
$\phi$ is to be interpreted as the expectation value of A when S is in the 
state $\phi$. 

A symplectic form is introduced on $\mathcal{A}$ and PBs (denoted as $\{, \}$) 
defined. 
A (Hermitian) element H of $\mathcal{A}$ is chosen as the 
Hamiltonian. To describe evolution of the system, time dependence may be put 
in observables or states (corresponding, respectively, to a Heisenberg or 
Schr$\ddot{o}$dinger type picture), the two being related as 
$\phi (t)(A) = \phi (A(t))$. The corresponding evolution equations are 
(called, respectively, the  Hamilton's equation and  
Liouville's equation of supmech)
\begin{eqnarray}
\frac{d A(t)}{dt} = \{ H,A(t)\} \equiv D_H (A(t) \\
\frac{d \phi(t)}{dt}(A) = \phi(t)(\{ H,A \}) \equiv (\tilde{D}_H \phi(t))(A) 
\end{eqnarray}
where $ \tilde{D}_H$ is the transpose of the operator $D_H$ (in the sense of 
an operator on the dual space of $\mathcal{A}$). Finite time evolutions may 
be written (in situations where the exponentials below can be properly 
defined) as 
\begin{eqnarray}
A(t) = exp[(t-t_0)D_H]A(t_0); \hspace{.15in} 
\phi(t) = exp[(t-t_0)\tilde{D}_H]\phi(t_0).
\end{eqnarray}

\noindent 
Note. (i) For PBs, we adopt Woodhouse's [14] conventions; these differ 
from most mechanics texts by a minus sign. \\
(ii) In Ref. [9], the formalism is developed in a superalgebraic setting 
(taking $\mathcal{A}$ to be a superalgebra) to achieve a unified treatment 
of bosonic and  fermionic objects. Here the non-super version is adequate. 

Classical mechanics and traditional QM can be realised as subdisciplines of 
supmech. In the former case, we have $ \mathcal{A} = C^{\infty}(\Gamma)$, the 
usual PBs and Eqs. (3) and (4) become the usual Hamilton and Liouvile 
equations. In the latter case, $\mathcal{A}$ is an algebra of operators on the 
quantum mechanical Hilbert space $\mathcal{H}$ of the system (defined on 
a suitable dense domain in $\mathcal{H}$), the PBs are 
$ \{ A, B \}_Q \equiv (-i \hbar)^{-1}(AB - BA)$; Equations (3,4) become the 
Heisenberg and von Neumann equations (the latter is equivalent to the 
Schr$\ddot{o}$dinger equation when $\phi$ is a pure state).

\noindent
\textbf{3.} Interactions of two systems $S_1$ and $S_2$ [either or both of 
which can be commutative (classical) or noncommutative (quantum)] can  be 
treated in supmech by taking, for the system ($S_1 + S_2$), the algebra 
$ \mathcal{A} = \mathcal{A}_1 \otimes \mathcal{A}_2$ (the tensor product of 
the algebras $\mathcal{A}_1$ and $\mathcal{A}_2$ corresponding to $S_1$ and 
$S_2$; elements of $\mathcal{A}$ are finite sums $\sum_i A_i \otimes B_i $
with $A_i \in \mathcal{A}_1$ and $B_i \in \mathcal{A}_2$.) States of 
$\mathcal{A}$ are of the form $\phi_1 \times \phi_2$ (and weighted sums of 
such states) where $\phi_1$ and $\phi_2$ are states of $S_1$ and $S_2$ 
respectively, such that $ (\phi_1 \times \phi_2)(A \otimes B) = 
\phi_1(A) \phi_2(B)$. Denoting the PBs on $\mathcal{A}_1$ and 
$ \mathcal{A}_2$ by $\{, \}_1$ and $\{, \}_2$, the PB on $\mathcal{A}$ is 
given by [9] 
\begin{eqnarray}
\{ A \otimes B, C \otimes D \} = \{A, C \}_1 \otimes 
\frac{BD + DB}{2} + \frac{AC + CA}{2} \otimes \{ B, D \}_2.
\end{eqnarray}
Eq. (6) is easily verified for the cases when (i) both algebras are those of 
functions on phase spaces and (ii) both are matrix algebras.

Denoting by $H_1$ and $H_2$ the Hamiltonians for $S_1$ and $S_2$, and by 
$I_1$ and $I_2$ the unit elements in $\mathcal{A}_1$ and $\mathcal{A}_2$, 
the Hamiltonian for $(S_1 + S_2)$ is taken to be of the form 
\begin{eqnarray}
H = H_1 \otimes I_2 + I_1 \otimes H_2 + H_{int} 
\end{eqnarray}
where the interaction Hamiltonian is generally of the form (absorbing the 
coupling constants in algebra elements)
\begin{eqnarray}
H_{int} = \sum_{i = 1}^{n} F_i \otimes G_i. 
\end{eqnarray}
A typical 
observable $ A(t) \otimes B(t)$ of ($ S_1 + S_2 $) evolves as
\begin{eqnarray}
\frac{d}{dt} [A(t) \otimes B(t)] = \{ H_1, A(t) \}_1 \otimes B(t) + 
A(t) \otimes \{ H_2, B(t) \}_2  \nonumber \\
+ \{ H_{int}, A(t) \otimes B(t) \};
\end{eqnarray}
 restriction to observables in $S_1$ ($S_2$) can be achieved by taking 
$ B = I_2 \ (A = I_1)$.

\noindent
\textbf{4.} A formalism like supmech which has classical and quantum 
mechanics as straightforward subdisciplines has (at least) two good potential  
applications : (i) a consistent description of the interaction of quantum 
matter and classical gravity and (ii) measurements in QM. About (i), we 
only point out that the two (quantum matter and classical gravity ) can be 
treated as supmech Hamiltonian systems and that the interaction can be 
treated as above, leaving the details for a future publication. Item (ii) is 
treated below.

\noindent
\textbf{5.} We shall now treat the (S + A) system of section (1) in supmech. 
The description of S and A as a quantum and a classical system in supmech are as indicated at the end of section (2). Note that an observable $ A \otimes f$ 
of (S + A) can also be written as fA. We shall freely employ the two notations. 

Two important points about the apparatus A are : 

\noindent
(i) the observations relating to 
it are restricted to the pointer positions $M_j$ which correspond to disjoint 
domains $D_j$ in the phase space $\Gamma_A$ of A; 

\noindent
(ii) different pointer positions are macroscopically distinguishable. 

\noindent
A general pointer observable 
for A is of the form 
\begin{eqnarray}
J = \sum_j b_j \chi_{D_j}
\end{eqnarray}
where the $\chi_D$ is the characteristic function for the domain D and 
$b_j$s are nonzero real numbers such that $b_j \neq b_k$ for $j \neq k$. The 
pointer state $\phi^{(A)}_j$ corresponding to the pointer position $M_j$ is 
represented by the phae space density function $ \rho_j = V(D_j)^{-1} 
\chi_{D_j}$ where V(D) is the phase space volume of the domain D. [Note. The 
appearance, in J above, of non-smooth characteristic functions is not an 
essential complication; they can be represented arbitrarily closely by 
appropriately chosen smooth functions.]

We shall take $H_{int} = F \otimes K$ (absorbing the coupling constant in K) 
where F is the measured quantum obsevable and K is a suitably chosen apparatus 
observable (it is expected to have nonzero PB with the relevant pointer 
observable). We shall make the usual assumption that, 
during the measurement interaction, $H_{int}$ is the dominant part of the 
Hamiltonian ($H \simeq H_{int}$). The object replacing the unitary operator 
U of section (1) is the measurement interaction operator $ M = exp[\tau 
\tilde{D}_H]$ where $ \tau = t_f - t_i$ is the time interval of measurement. 
It represents a canonical transformation on the states of the system (S+A). 
The analogue of the equation describing the action of U above is [denoting 
the `ready' state of the apparatus by $\phi_0^{(A)}$]   
\begin{eqnarray}
M (|\psi_j><\psi_j| \times \phi^{(A)}_0) = |\psi_j><\psi_j| \times 
\phi^{(A)}_j.
\end{eqnarray}

Note that the `ready' state may or may not correspond to one of the pointer 
readings. (Examples : It does in voltage type measurements; it does not in 
the Stern-Gerlach experiment with spin half particles.) For assignment of a 
$\Gamma$-domain to the `ready' state (which is, from the theoretical point 
of view, very important; see below) its proper interpretation is `not being in 
any of the (other) pointer states. Accordingly, we shall assign it the domain 
\begin{eqnarray}
\tilde{D}_0 \equiv \Gamma - \cup_{j \neq 0} D_j
\end{eqnarray}
where the condition $j \neq 0$ is to be ignored when the ready state is not 
a pointer state.

Since the apparatus is classical, its state at any given time will be in 
any one of the domains $D_j$. With the system in the initial superposition 
state $|\psi>$ as above, 
the initial and final (S + A)-states 
in the supmech description are
\begin{eqnarray}
\Phi_{in} = |\psi><\psi| \times \phi^{(A)}_0 ; \ \ \Phi_f = M(\Phi_{in}).
\end{eqnarray}
 The final state 
expected on applying von Neumann reduction is [the analogue of the state (2) 
above]
\begin{eqnarray}
\Phi^{\prime}_f & = & \sum_j |c_j|^2 [ |\psi_j><\psi_j| \times \phi^{(A)}_j]
                                            \nonumber \\
		& = & M \left( \sum_j|c_j|^2 [ |\psi_j><\psi_j| \times 
		      \phi_0^{(A)}] \right)
\end{eqnarray}
where we have used the fact that a canonical transformation on states 
preserves convex combinations.

Recalling the apparatus feature (i) above, the most suggestive  way the 
state (2) or its 
supmech analogue (14) appear to  make sense is as the effective final states 
of an observationally constrained system.
We shall show that, for a general system observable A and a pointer observable 
J, the expectation values of the (S+A)-observable $A \otimes J$ in the states 
$\Phi_f$ and $\Phi^{\prime}_f$ are equal (to a very good approximation). 
We, therefore, look for the vanishing of the quantity 
\begin{eqnarray*}
W \equiv (\Phi_f - \Phi_f^{\prime}) (A \otimes J) = M(R)(A \otimes J)
\end{eqnarray*}
where 
\begin{eqnarray*}
R = [\sum_{j \neq k} c_k^* c_j |\psi_j><\psi_k|] \times \phi_0^{(A)}.
\end{eqnarray*}
[Note that R is not an (S +A) state; here M has been implicitly extended by 
linearity to the dual space of the algebra $\mathcal{A}_S \otimes 
\mathcal{A}_A$.] 
Transposing the M-operation to the observables, we have 
\begin{eqnarray}
W & = & R [exp(\tau D_H) (A \otimes J)] \nonumber \\
  & = & \frac{1}{V(\tilde{D}_0)} \int_{\tilde{D}_0} d \Gamma \sum_{j \neq k} c_k^* c_j 
        < \psi_k|exp(\tau D_H)(A \otimes J)|\psi_j>
\end{eqnarray}
where $d \Gamma$ is the phase space volume element. Now 
\begin{eqnarray}
D_H(A \otimes J) & = & \{ F \otimes K, A \otimes J \} \nonumber \\
                 & = & (-i \hbar)^{-1}KJ [F,A] + \{ K,J \}_{cl} 
		       \frac{FA + AF}{2}.
\end{eqnarray}	
Writing the second term on the right in Eq.(16) as $Q(A \otimes J)$, we can 
write, in Eq.(15), $ D_H = bI + Q $ with $ b = (-i\hbar)^{-1} 
(\lambda_j - \lambda_k)$. This gives 
\begin{eqnarray}
W = \frac{1}{V(\tilde{D}_0)} \int_{\tilde{D}_0} d \Gamma \sum_{j \neq k} 
c_k^* c_j 
exp[\frac{i}{\hbar} (\lambda_k - \lambda_j)K \tau]. \nonumber \\
.<\psi_k|exp(\tau Q)(A \otimes J)|\psi_j>.
\end{eqnarray}

We shall now argue that $ |\eta| > > \hbar$ where $ \eta \equiv 
(\lambda_k - \lambda_j)K \tau $. (This is not obvious; in the case of spin 
measurements, for example, the $\lambda$s are scalar multiples of $\hbar$.)
To ensure the apparatus feature (ii) above, a reasonable criterion is 
$ |\Delta E \Delta t| > > \hbar $ where $ \delta t = \tau$ and $ \Delta E $ 
the difference between energy values corresponding to the apparatus locations 
in two different domains $ D_j$ and $D_k$ in $\Gamma$. Recalling that 
$ H \simeq H_{int}$, we have $ \Delta E 
 \simeq (\lambda_k - \lambda_j)K $, completing the argument.

The large fluctuations implied by $ |\eta| > > \hbar$ wipe out the 
integral giving $ W \simeq 0 $ [which should be interpreted as W = 0 `for 
all practical purposes' (FAPP)] as desired. 

What has been proved above is this : Given the observable F and the 
descriptions of the system and apparatus as above and the ideal 
measurement condition that, 
if the system is initially in an eigenstate of F, the measurement outcome is 
the corresponding eigenvalue ( and the system is left in the same state 
at the end of the measurement), the following is true FAPP :
when the system is initially in a superposition  state $\psi$ as above, 
the outcome in every individual measurement  is an eigenvalue of F (and the 
system 
at the end of the measurement is in the corresponding eigenstate); the 
probability of the eigenvalue $\lambda_j$ appearing as the outcome is 
$|c_j|^2$. 

\noindent
\textbf{6}. Remarks. (i) The derivation above also makes it clear as to the 
sense in which the projection rule in QM should be understood -- it is a 
prescription  for obtaining the effective final state of the observationally 
constrained (S+A) system. 

\noindent
(ii) In the formalism above, the environmental effects may be understood 
as taken implicitly into consideration. [One need only say that the system A 
above represents `apparatus and environment' (the phase space of the 
environment now understood as part of the domain $\tilde{D}_0$); the whole 
argument goes through -- without any extra efort.] In fact, even when the 
external environment is not considered, the integration over $\tilde{D}0$ 
in Equations (15) and (17) is essentially the averaging over the internal 
environment [4] of the apparatus. 

\noindent 
(iii) In the calculations after Eq.(15), the explicit form (10) of J was 
not used. Then why talk of an obsrvationally constrained system ? The 
justification is that the feature (i) of the apparatus was used in the 
identification of the domain $\tilde{D}_0$ which played an important role 
in the final calculations.

\noindent
(iv) Main message : The Hilbert space framework is not adequate for the 
development of QM (this has already been noted in the context of quantum field 
theory [15]); the proper framework is provided by supmech.

\vspace{.15in}
\noindent
\textbf{Acknowledgements}

\noindent
The author would like to thank R. Sridharan and V. Balaji for helpful 
discussions, to Chennai Mathematical Institute for support and the 
Institute of Mathematical Sciences, Chennai for library and other facilities.

\begin{description}
\item[{[1]}] J. von Neumann, \textsl{Mathematical Foundations of Quantum 
Mechanics}, Princeton University Press (1955).
\item[{[2]}] J.A. Wheeler and W.H. Zurek, \textsl{Quantum Theory and 
Measurement}, Princeton University Press (1983).
\item[{[3]}] R. Omnes, \textsl{The Interpretation of Quantum Mechanics}, 
Princeton University Press (1994).
\item[{[4]}] Tulsi Dass, `Measurements and Decoherence', quant-ph/0505070.
\item[{[5]}] W. H. Zurek, `Decoherence, Einselection and Quantum Origin of the 
Classical', Rev. Mod. Phys.  \textbf{75} (2003), p.715.
\item[{[6]}] J. Bub, \textsl{Interpreting the Quantum World}, Cambridge 
University Press (1997). 
\item[{[7]}] S.L. Adler, `Why decoherence has not solved the measurement 
problem: a response to P.W. Anderson', quant-ph/0112095.
\item[{[8]}] W.H. Zurek, `Decoherence and the transition from quantum to 
classical--revisited', quant-ph/0306072.
\item[{[9]}] Tulsi Dass, `Supmech : a Unified Symplectic View of Physics',
to be published.
\item[{[10]}] Tulsi Dass, `Towards an Autonomous Formalism for Quantum 
Mechanics', quant-ph/0207104.
\item[{[11]}] M. Dubois-Violette, R. Kerner and J.Madore, J. Math. Phys. 
\textbf{31} (1994), p.316.
\item[{[12]}] M. Dubois-Violette, `Noncommutative Differential Geometry, 
Quantum Mechanics and Gauge Theory' in \textsl{Lecture Notes in Physics, 
vol 375}, Springer, Berlin (1991).
\item[{[13]}] M. Dubois-Violette, `Some Aspects of Noncommutative Differential 
Geometry', q-alg/9511027.
\item[{[14]}] N. Woodhouse, \textsl{Geometric Quantization}, Clarendon Press 
Oxford (1980).
\item[{[15]}] G.E. Emch, \textsl{Algebraic Methods in Statistical Mechanics 
and Quantum Field Theory}, Wiley, New York (1972).

\end{description}
\end{sloppypar}

\end{document}